# Reflections on the Clinical Acceptance of Artificial Intelligence

Jens Schneider and Marco Agus

**Abstract** In this chapter, we reflect on the use of Artificial Intelligence (AI) and its acceptance in clinical environments. We develop a general view of hindrances for clinical acceptance in the form of a pipeline model combining AI and clinical practise. We then link each challenge to the relevant stage in the pipeline and discuss the necessary requirements in order to overcome each challenge. We complement this discussion with an overview of opportunities for AI, which we currently see at the periphery of clinical workflows.

## 1 Introduction

To say that Artificial Intelligence (AI) has matured enough over the last decades to be of practical significance would be a clear understatement. As of writing in May 2020, AI is generally regarded as disruptive technology, creating its own job profiles (e.g., Data Scientist), impacting a wide range of industries (e.g., the automotive industry in the form of autonomous vehicles), and spawning new academic programs to cope with the ever increasing demand for skilled man power in the field. Although the Business Insider magazine in 2017 [4] listed Healthcare as one out of 9 industries that are being transformed by AI around the world, we continue to see what is best described as reluctance in the healthcare sector to fully embrace AI to the extent that other industries already have. But why is that so? Certainly, clinical environments lend themselves to a degree of conservatism, and, one might add, for the better: Many new techniques are not necessarily battle-proven to the standards of clinical rigor.

———————————————

Jens Schneider
College of Science and Engineering, Hamad Bin Khalifa University, Doha, Qatar.
e-mail: jeschneider@hbku.edu.qa

Marco Agus
College of Science and Engineering, Hamad Bin Khalifa University, Doha, Qatar.
e-mail: magus@hbku.edu.qa

But that alone can hardly explain why, in our opinion, the healthcare sector is slow to adopt to AI-powered technology. To some extent, we agree with the findings of a recent survey on AI-based technology and its use to fight COVID-19 [3] that many AI technologies are not yet mature for clinical deployment. However, we believe that this is not a fundamental flaw of the technology itself, but rather the relative lack of well-conducted clinical studies [19]. In this chapter, we will therefore analyze reasons for this perceived technological immaturity and factors that hinder clinical acceptance of AI. We will further discuss how these challenges can be overcome, and discuss which types of applications and technologies are more likely to gain acceptance more quickly than others.

## 1.1 AI Nomenclature

This chapter is written from a data science perspective, but targeted at a much broader audience. In an attempt to make this chapter more self-contained, we have therefore taken the liberty to briefly discuss AI nomenclature in this section.

AI can solve different problems, such as *classification* (e.g., given chest x-ray images, does a patient have pneumonia and if so, is it caused by virus or bacteria), various *localization* tasks (e.g., segmentation of structures), processing and interpreting natural languages, etc. The general mechanism is that there is an *architecture* (a term, loosely speaking, relating to the combination of mathematical building blocks, not unlike a popular Danish plastic toy for children) which is home to a plethora of unknown variables. *Training* is then the process to iteratively update these variables by feeding a subset of often annotated data (*training data*) into a robust, numerical optimizer. At the beginning of the training process, variables are initialized with random numbers. The answer to any given query (such as: does this x-ray indicate pneumonia?) will therefore be quite erratic. The deviation between the answer of this mathematical prediction machinery and the known answer or *label* is called the *loss*. Methods using such labels are generally called *supervised* learning—*unsupervised* methods that do not use labelled data are not discussed in this chapter. However, AI is not concerned with memorizing the training data, it is concerned with predicting some aspect of future data samples (class, segments, ) it has not encountered before. Therefore, an independent, second data set, the *validation data* is used to ensure the AI can make predictions for unknown data samples. The underlying as-sumption is that of *statistical generalization*—if the training set is large enough, the insights gained from minimizing the loss generalize to make accurate statements about unknown data from the validation set. The combination of architecture and the numbers learned are called the *model*. The model is the very heart of any AI method, and training good models requires large amounts of data and computational power.

## 2 AI in Clinical Environments

The aforementioned Business Insider article [4] highlights a UK-based company called Babylon Health that developed a chat-bot based on AI technology such as Natural Language Processing (NLP) that is used by the UK's National Health Service (NHS) as a contact front-end for patients. Along a similar venue, more specialized chat bots targeting mental health patients in the Arab world have been proposed [5, 6], particularly addressing that language barriers may exacerbate the sensitive issue at hand. Albeit not exactly set in a clinical environment, we will analyze in the following why this particular technology seems to have gained acceptance whereas others do not have (yet).

Another application where AI is likely to be integrated into clinical workflows is skin lesion diagnosis using commodity hardware. Driven by the image data made publicly available through the International Skin Imaging Collaboration (ISIC) [7], the international AI community has picked up the challenge and delivered solutions that achieve accuracy scores above 0.866 (best balanced multiclass accuracy in the ISIC 2018 competition, e.g., [15])——high enough to be of clinical value. In a nut-shell, any person with a handheld device such as a smartphone or tablet equipped with a special yet inexpensive clip-on zoom lens can self-diagnose moles and be referred to a MD if necessary. We would like to make two observations regarding this application that we deem important. Firstly, we see the technology developed in this context not so much being used by seasoned dermatologists, but rather by general physicians. The reason is that this technology, for the first time, offers them a way to obtain a second opinion inexpensively, and quickly. It may, therefore, well enter the routine of general physicians to scan their patients, and, if need be, refer them to specialists. Secondly, the innovation behind this technology was driven by rather minuscule incentives (a $4,000 cash price for the best entry every year), very similar in style to the now famous DARPA Grand Challenge of 2005 [2], in which a moderate cash prize (in comparison of the US army's monetary efforts up to this point) incentivized the academic, engineering, and technology crowd. It is not only our firm belief that these DARPA challenges had a significant role in kick-starting what has now become the driver-less car industry.

On the other hand, the first author of this chapter briefly worked as an intern in a German university spin-off company developing a computer-based pre-operative planning system for full knee and hip replacements, more than one and a half decades ago. At that time, we were developing an expert system that would recom-mend position and orientation of implants based on CT scans and a simulation of the ligaments (for the knee replacement). Although similar systems eventually found use at university-affiliated hospitals , certification and registration as a medical device is usually a long and tedious process under most legislations.

We believe that the difference is, mainly, where innovation happens. In both im-mediate success stories (chatbot and skin lesion diagnosis), innovation happens at "the edge" of the clinical environment, a space that is agile and can pivot quickly. In contrast, the chances for latest technology ("latest" from a research perspective) to be used in "life and death" scenarios is rather scant. The skin lesion system recommends either to see or not to see a physician, and, as long as false negatives can be minimized below the probability of patients not undergoing regular cancer screenings, creates tangible value in a clinical setting. The chatbot developed by Babylon Health essen-tially streamlines and augments hospital receptions, which may boost productivity while not critically affecting patient treatment. In contrast, full knee or hip pros-thetic replacement comprises surgery, and ill-advised prosthetic placement has the potential to affect patients' lifestyle for years to come, including frequent follow-up surgeries. It is, therefore, understandable that exhaustive documentation and stud-ies are normally required for clinical certification and registration of such technology.

The fundamental question, however, is: Do these observations generalize? What are the lessons learned from such anecdotal evidence and how can we further for-malize the challenges and opportunities faced in the clinical acceptance of AI-based technology? To understand the problem better, let us sketch how we, from the data science perspective, see the flow from medical data (and all AI is oh so very data hungry) to clinical acceptance, linked to factors potentially hindering this very flow.

## 3 Challenges & Opportunities

Figure 1 summarizes our attempt at a general view of the hindrances affecting clini-cally approved AI-based technology. In the remainder of this section, will follow the pipeline from data to full clinical acceptance in this figure, we will discuss the chal-lenges associated with each stage and necessary steps to overcome the hindrances.

### 3.1 Data Repositories

As the scenario of skin lesion diagnosis helped motivate, we believe that the first step is to create an inter-institutional data repository. Data is the source of all AI, and the data of one hospital rarely suffices. This, in turn raises questions of data sovereignty and privacy, requiring data to be anonymized in the very least. More subtly, many researchers in the biomedical field grow rather fond of their data, since it represents a substantial investment of time and money on their behalf. Data sovereignty issues may thus potentially be exacerbated by a certain degree of academic mistrust within the community. Another issue obstructing the creation of such repositories are a plethora of incompatible, proprietary, or poorly documented data formats. In the context of the COVID-19 pandemic, one survey explicitly called for the public release of existing case studies [19], finding that this is far from the current practise. We second this and note that in the shadow of the COVID-

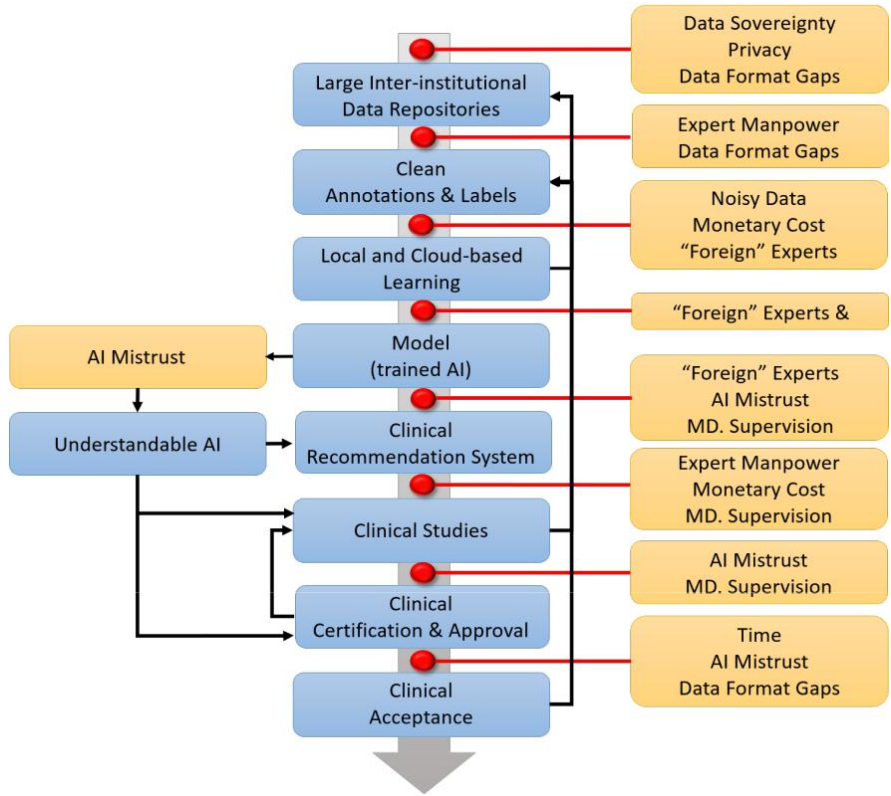

**Fig. 1** Flow from medical data to clinically accepted AI technologies and risks.

19 pandemic, we are approaching 5 million cases with over 300,000 deaths as of writing [18], yet, according to our own research, very few studies seem to have access to large data (e.g., only 1 out of the 15 studies included in a recent survey [19] contained more than 1,000 data samples). Furthermore, big companies are collecting their customers' data and may not be willing to share such data on the grounds that this data is key to their own AI endeavours. Regarding clinical or governmental data, academic collaborations can help democratize such data and government incentives, Digital or Open Government approaches [11] likewise seem adequate mechanism to ensure that the research community has access to large, unbiased, balanced, and diverse data sets. Regarding commercial data, however, the data often represents the commercial edge of tech companies, and the only remedy seems to be careful legislative interventions in which the greater good has to be constantly evaluated against the companies' rights to maintain independent business operations.

## 3.2 Clean Annotation & Labels

Once a repository is created and can be shared with AI researchers, the tedious process of cleaning as well as annotating and labelling the data sets in to provide AIs with a concrete goal for their training. While data scientists may aid in cleaning the data, annotations and labels have to be provided by medical experts. This requires substantial manpower on both the AI and the clinical side. In this and later stages, we assume the view of the clinical environment and view ourselves, the data scientists as "foreign" experts. This shift of view is an exercise mostly motivated by the fact that we commonly see a significant language disconnect or even an outright language barrier between data sciences and health care professionals. This hinders communication and coordination between the two fields. Despite new programs of study targeted at bridging this gap, the authors are not sure if this issue can be successfully overcome in the future, as both fields, and in particular AI, are continuously developing.

## 3.3 Local and Cloud-based Learning

Finally, we have arrived at the point where a model can be trained. Moderate incentives provided, a clinical problem or challenge could now be outsourced to an army of data scientists and researchers. However, the first hurdle encountered in this stage is that the data is still too noisy to be of practical use. Part of it may actually be attributed to the gap between the medical environment and "foreign" experts. Both the medical profession and data scientists tend to groom their data meticulously. However, the methods could not be more different. Data science models data in clean, quantitative terms, striving to reduce information theoretical redundancy and favoring such structure that enables computation on the data. In contrast, medical data is optimized for human understanding. In a sense, it is richer, more diverse, and, regrettably, tends to be more prone to noise. The reason is that apart from quan-titative data, medical professionals also record qualitative data, such as: How does the patient's health improve? What is the pain level? Is the progress better or worse than cases previously encountered by the physician? All these data points are hard to put in numbers, and, physicians consequently prefer to describe their qualitative findings in natural language (with high redundancy), making notes in one of many information systems (data format gap). Consequently, this learning stage may back-track into the cleaning and data repository steps. Once the data is sufficiently clean, the last hurdle is the monetary cost involved in training AI, sometimes for months on end and on high-end computing hardware. Since much of today's training happens in the cloud, it is important to note that privacy and data sovereignty issues have to be solved before even approaching this stage.

### 3.4 Model (trained AI)

In this stage, a trained AI is at the fingertips of medical experts and data scientists. Let us assume for now that it does something useful, with an accuracy that is well above uninformed guess work. But how does it actually do it? For sure, we can assume that if we present the AI with the same case it has seen before, it will do its job and return the answer we provided for it in the form of labels. But that is just mere data retrieval; how can we know that the AI is right for a new case, that it *generalizes*? The nagging sensation that the freshly trained AI might be resorting to black magic altogether, called *AI mistrust*, is a very serious hindrance for all stages that follow and a very valid concern [13]. A new research direction called understandable AI spawned recently to analyze how the AI actually generalizes from the cases presented to new ones. Researchers in this field try to disassemble and visualize deep networks in order to understand what exactly it is each layer in a deep network does and if the overall network can be trusted. We believe that this field is crucial for the penultimate clinical acceptance of AI, and we would not be surprised if questions regarding explainability and plausibility of the AI as well as the ethical implications of AI were to become mandatory in the clinical certification and approval process in the near future [14].

### 3.5 Clinical Recommendation System

Assume you are a data scientist and you have successfully overcome all the hurdles so far. You have trained a model from clinically relevant data and you have built a recommendation system that uses AI. You would, of course, like to advertise it as a panacea that automatically diagnoses a wide range of diseases or makes recommendations so profound that you fully expect it to replace squadrons of medical professionals within the next couple of months. But you are aware of the ethical implications and you suspect you have succumbed to hubris of the advanced kind. Therefore, you humbly state that a physician has to verify the result of your AI and has to decide its use in any clinical treatment. In short, your system advises a physician whereas the physician supervises your system. It therefore becomes a *tool*, with liability implications that we will discuss later in this chapter. This scenario is of value, to varying degrees, depending on the context. If the recommendation system operates at the non-critical edge of the healthcare system, it may end up adding significant value. If it is targeted at the core of the health care system (e.g., operation theaters where time is extremely valuable and experience is everything), it may never be accepted. We believe that the key to understanding this difference are two simple questions: How much time does the supervisor have / is willing to spend in order to get an advise? How much time will the advice potentially save or how much will it improve the patient's and doctor's life? Going back to the example of skin lesion diagnosis: The AI is a recommendation system that sits at the edge of the health care system. It does not cost much time or effort and simply provides a second opinion. The risk of not treating a range of skin cancers in a range of patients will most likely convince many non-dermatologists to use this system. Dermatologists may evaluate the technology based on its accuracy and the amount of time the AI saves for them.

## 3.6 Clinical Studies

The next stage are clinical studies. This stage requires expert manpower, time, money, and supervision through healthcare professionals. In this context, clinical studies should not only assess whether the AI is reliable and accurate, but they should also try to assess the ethical and social implications of using an AI. Are physicians and patients comfortable with taking advice of an AI? Will the AI be in direct contact with patients or only the professionals? What is the improvement in the professionals' workflows? Do they actively supervise the technology or do they grow fatigued and accept the advice uncritically?

## 3.7 Clinical Certification & Approval

This stage essentially takes AI from research to a medical product. Understandably, this step takes time and requires substantial involvement of physicians, lawyers, data scientists, etc. Since systems for clinical use have to be fully documented, again AI mistrust may be a hurdle that can be overcome by demonstrating that the AI is plausible and understandable. It is worth noting though that the time required to achieve approval generally translates into technology out-dating in the process. This is all the more true for AI-based solutions which are fueled by one of the fastest growing research fields at the moment.

## 3.8 Clinical Acceptance

We do not equate the mere fact that an AI has undergone certification and has been approved to clinical acceptance. We define clinical acceptance as:

*A significant portion of major hospitals has heard of a technology and either considers its use or is using that technology.*

Again, there are a wide range of reasons why an approved technology would not become accepted. For instance, the manufacturer of a product is unable to integrate the product into a hospital's workflow or IT infrastructure (data format gap), the hospital is not convinced about the benefits of AI-based products (AI mistrust), or the product is simply too new to be widely accepted.

## 3.9 Liability Risks

Even though AI is still far from being largely accepted in core clinical setups, we feel that it is important to mention potential issues related to legal liability in addition to the risks outlined in Figure 1. We consider this in some ways very similar to the liability issues posed by recent advancements in autonomous driving vehicles [8], or all artificially intelligent computer systems. The main general questions can be summarized as follows:

- Who is liable (AI developer vs. medical practitioner) ?
- Whether and when shall criminal or civil code be applied (e.g., for malpractice) ?
- Is AI is an instrument, a tool, a service, or more ?

The interaction between physicians who tend to see AI as tools and the AI developers boils down to the following conflict: neither party wants to accept liabilities. This is probably one of the reasons clinical approval processes are so involved. We believe that, in order to understand liability better, we have to distinguish between the following scenarios:

- Medical malpractice, in which physicians risk law suits from patients who feel that the physicians were negligent in their treatment.
- Uncritical and uncontrolled reliance on automatic diagnosis and treatment deci-sions.
- Technical malpractice, in which AI achieves clinical certification despite un-documented, untested, or overlooked erratic behaviour that can lead to wrong decisions.

While it seems that clinicians tend to blame responsibility for wrong decisions on system providers, the same system provides tend to contractually exclude such lia-bilities. Therein lies a dilemma that will generate many future law controversies [1]. To this end, the European Commission has very recently faced this delicate issue and published a report that looks at whether the existing liability regimes are sufficient for the purposes of attributing liability in relation to highly complex tools such as AI and emerging technologies [16, 17]. The report highlights that a person using a technology that has a certain degree of autonomy should not be less accountable for harm caused than if said harm had been caused by a human aide. However, as AI technology will continue to evolve, we believe that the existing legislative framework for tort and product liability will need to be adapted accordingly. In the meantime, all stakeholders will need to assess whether or not they are sufficiently protected against liability risks arising from usage of AI, be it as operators, users, or manufacturers. This could be by way of contractual arrangements (e.g. warranties and indemnities), or by taking out appropriate insurance coverage. From our point of view, we do not feel in the position of expressing an opinion on these delicate issues. However, we wanted to point out that we firmly believe that, when an AI-based decision sup-port system is planned to be used in a clinical environment, a deep discussion is necessary at different levels to define specific responsibilities, starting from design specifications, to product implementation, and usage instructions and limitations.

## 4 Conclusion

Given all these considerations, we believe that the greatest opportunities for AI are currently on the edge of the healthcare system. Here, AI-powered applications can bypass full medical certification since they are not yet mission critical. Challenges of data sovereignty and privacy remain, but AI targeted at this segment may also well pave the way for a wider base of AI acceptance in the patient population. Consider-ing that AI and data science have become much more accessible during the last few years due to major advances in API design, another way to bridge the gap between data science and medical profession would be to further abstract typical algorithmic tasks, like the training and validation process, in a way to enable physicians with no coding experience to build automated deep learning models that might once have been out of reach [13]. This may be further assisted by data scientists already on pay roll in large hospitals. Being exposed to the way AIs are designed might also help in reducing AI mistrust.

The largest challenge we see is the need to democratize data. The largest and most valuable source of data in healthcare arguably comes from Electronic Medical / Health Records (EMRs / EHRs). However, clinician satisfaction with EMRs is still very low, with regards to completeness and quality of data entries [9]. This is made worse by inter-operability issues between different providers. At the same time, EMRs raise an interesting question: who owns the data in the EMR? Clearly, patients contribute their private data to an EMR, so there is at least partial ownership. However, also the physicians contribute to the EMR in the form of diagnoses, prescriptions, etc. Should, therefore, hospitals also assume partial ownership of the EMR? We believe not, since they were paid for their service and, therefore become consultants to their patients. We cannot rule out that this assessment might be wishful thinking from a data scientist perspective though, since it would imply that patients can voluntarily disclose all the data collected in their EMRs and all their history stored at hospitals into public or third party repositories, such as the now defunct Google Health or likewise defunct Microsoft HealthVault [10]. Questions such as these, of data ownership, liability distribution, responsibility and permission to use are at the very core of realizing the full potential of AI across health systems. In general, we can expect that the prevalent scenario for data infrastructure development will depend more on the socio-economic context of the health system in question rather than on technology. In the current status, the potential of AI is sufficiently highlighted, but in reality, health systems are faced with a dilemma: to significantly reduce the enthusiasm regarding the potential of AI in everyday clinical practice, or to resolve issues of data ownership, liability and trust, and to invest in the data infrastructure to realize it [12]. Such considerations will, eventually, tie back to and define AI ethics, a field currently emerging in academia.